\documentclass[10pt,a4paper,twocolumn,english,pra,aps,showpacs,floatfix,groupedaddress,superscriptaddress]{revtex4-1}

\usepackage{graphicx}
\usepackage{epsfig}
\usepackage[english]{babel}
\usepackage{amsmath}
\usepackage{amssymb}
\usepackage{amsfonts}
\usepackage{longtable}
\usepackage{dcolumn}
\usepackage{bm}
\usepackage{bbm}
\usepackage{color,array}
\usepackage{colortbl}

\usepackage{soul}

\newcommand{\uuline}[1]{\underline{\underline{#1}}}
\newcommand{\uline}[1]{\underline{#1}}

\begin{document}
\title{Effects of Interactions on Dynamic Correlations of Hard-Core Bosons at Finite Temperatures}

\author{Benedikt Fauseweh}
\email{benedikt.fauseweh@tu-dortmund.de}
\affiliation{Lehrstuhl f\"{u}r Theoretische Physik I, Technische Universit\"at Dortmund, 
Otto-Hahn Stra\ss{}e 4, 44221 Dortmund, Germany}

\author{G\"otz S.\ Uhrig}
\email{goetz.uhrig@tu-dortmund.de}
\affiliation{Lehrstuhl f\"{u}r Theoretische Physik I, Technische Universit\"at Dortmund, 
Otto-Hahn Stra\ss{}e 4, 44221 Dortmund, Germany}

\date{\rm\today}

\begin{abstract}
We investigate how dynamic correlations of hard-core bosonic excitation at finite temperature are affected 
by additional interactions besides the hard-core repulsion which prevents them from 
occupying the same site. We focus especially on dimerized spin systems, where these additional 
interactions between the elementary excitations, triplons,
lead to the formation of bound states, relevant for the correct description of 
scattering processes. In order to include these effects quantitatively we extend the 
previously developed Br\"uckner approach to include also nearest-neighbor (NN) and 
next-nearest neighbor (NNN) interactions correctly in a low-temperature expansion.
This leads to the extension of the scalar Bethe-Salpeter equation to a matrix-valued 
equation.  Exemplarily, we consider the Heisenberg spin ladder to illustrate the significance 
of the additional interactions on the spectral functions at finite temperature 
which are proportional to inelastic neutron scattering rates.
\end{abstract}


\maketitle

\section{Introduction}

Computing dynamic correlations in spin systems is one of the main tasks in order 
to understand the physics in real quantum magnets. Many exotic ground states without 
long-range order, for instance spin liquids, can be identified in
 neutron scattering experiments by their specific excitation spectra \cite{balen10}.
From an experimental point of view, thermal fluctuations often smear out the characteristic
signatures in momentum and frequency space making clear statements difficult 
\cite{paddi17}. This calls for theoretical predictions extended to finite
temperatures in order to directly compare with experiments. This goal, however, often proves 
challenging because at finite temperatures the full trace over the Hilbert 
space has to be taken into account, i.e., the complete Hilbert space contributes. 
Especially interactions between excitations 
can change the energy landscape significantly by means of  bound states or 
long-range entanglement.

Recently it was shown, that the Heisenberg ladder with strong inter-rung frustration 
is such an extreme case \cite{honec16}. This model exhibits bound states of the elementary
triplon excitations which exist even below the single-triplon gap. 
Interestingly, these bound states are hidden in the 
observables accessible by inelastic neutron scattering at zero temperature. At 
finite temperatures, however, the bound states acquire finite weight and can even 
dominate the spectrum. Thus, the low-energy physics is best described by strongly interacting
and entangled triplons. This analysis shows that the interactions between 
the elementary excitations can play a crucial role in the dynamics of spin systems
at finite temperature.  
 
On the methodical side, there exists a variety of methods to compute the dynamical 
response of a spin system at finite temperatures 
\cite{fabri97a,mikes06,essle08,james08,essle09,goetz10,exius10,lake13,jense14,tiege14,becke17}. 
In previous studies \cite{fause14,fause15},
we established an analytical method to calculate  correlation functions at finite temperature
based  on the Br\"uckner approach, first introduced  in nuclear physics and later
transferred to solid state physics \cite{kotov98}. 
It was gauged against exact data obtained from the Jordan-Wigner mapping to 
interaction-free fermions \cite{fause14}.
In contrast to most previous studies, the 
Br\"uckner approach has the asset that it is not restricted to one dimension or small
system size, but in return it relies on a small parameter, namely $\exp(-\beta\Delta)$
where $\beta$ is the inverse temperature and $\Delta$ the energy gap, so that it is 
particularly reliable at low temperatures.
It is important to dispose of methods which are applicable for all dimensions
because the thermal broadening of hard-core bosonic line shapes is observed
also in three dimensions \cite{quint12}. A good description has been obtained
by an expansion in the inverse coordination number \cite{jense14}. This approach, however,
is not justified in low dimensions. So the Br\"uckner approach is the only
one which is conceptually applicable in arbitrary dimension. 

The basic idea is to expand the single-particle Green function in 
terms of interaction diagrams and to keep only those diagrams which contribute in 
leading non-trivial order, i.e., in $\exp(-\beta\Delta)$.
These are the ladder diagrams in the self-energy of the single-particle propagator.
For gapped spin systems local excitations generically obey a 
hard-core constraint due to the limited size of the local Hilbert space. In order to be able 
to apply bosonic perturbation theory this hard-core constraint is incorporated as 
on-site infinite repulsion $U\to\infty$. 
The Br\"uckner approach was applied to quantitatively explain the
experimental data for two one-dimensional (1D) materials \cite{klyus16, fause16}.
It was also applied to predict such data in a two-dimensional (2D) material \cite{strei15}. 

So far, only the hard-core repulsion was taken into account in calculations
of the low-temperature spectral functions based on the Br\"uckner approach. 
Omnipresent additional interactions were included only on a mean-field level 
\cite{strei15,klyus16,fause16}. 
This appeared justified by the dominating strength of the diverging on-site repulsion 
in comparison to the additional \emph{finite} interactions. If, however, the additional 
interactions lead to a significant restructuring of the energy landscape, see discussion 
of binding phenomena above, a mean-field treatment 
is no longer justifiable. It is the main focus of this paper to solve this issue.

We derive how the Br\"uckner approach can be extended in order to include 
additional interactions summing all ladder diagrams. The extended approach 
correctly captures all scattering processes of two given particles including the formation
of bound states. The extension leads to a natural generalization of the scalar Bethe-Salpeter 
equation for the scattering amplitude to a matrix-valued Bethe-Salpeter equation
for the scattering matrix.

As a  testbed, we investigate the correlations at finite temperature for  
Heisenberg spin ladders. These systems feature triplons as elementary excitations 
which form bound and antibound states in the two-particle sector due to additional 
interactions. 
We investigate how these additional interactions influence the single-triplon 
spectral function at finite temperature. This paves the way to compute correlation 
functions in more complicated models and thus to explore the interplay of quantum 
interactions and thermal fluctuations in a broader sense.  
 
The article is set up as follows: In Sect.\ \ref{sec.model}, we introduce the hard-core
boson  model and the parametrization for the additional interaction. In Sect.\ \ref{sec.method},
we extend the Br\"uckner approach to additional interactions for hard-core bosons of
a single kind, i.e., for the single-flavor case. In Sect.\ \ref{sec.results}
an analysis of the approach for the Heisenberg ladder follows, for which
we extend it to several flavors as well. 
We conclude our article in Sect.\ \ref{sec.conclusion}.

\section{Model}
\label{sec.model}

Here we introduce the general hard-core boson model and 
discuss some of its properties. We consider a model with a single
kind of boson per site, i.e., a single flavor, in order to keep
the notation transparent. This setting can
be extended later on to several flavors.
The Hamiltonian of the system reads
\begin{align} 
\label{eq.model_ham}
H_0 &= E_0 + \sum\limits_{i,d} \left( h_d b_{i}^\dagger b_{i+d}^{\phantom\dagger} 
+ \mathrm{h.c.} \right) 
\nonumber  \\ &+
\hspace{-0.2cm}
\sum\limits_{i,  d_1,  d_2,  d_3} 
V_{d_1,d_2,d_3} b_{i}^\dagger b_{i+d_1}^\dagger  
b_{i+d_1+d_2}^{\phantom\dagger} b_{i+d_3}^{\phantom\dagger} 
+ \dots,
\end{align}
where $E_0$ is the ground state energy, $i,d,d_1,d_2,d_3$ are site indices and 
$b_{i,\alpha}^\dagger, b_{i,\alpha}$ are the hard-core bosonic creation and annihilation 
operators. We describe the approach for the one-dimensional case explicitly, but all 
definitions and equations can be implement straightforwardly in higher dimensions as well. 
The dispersion of the excitations is given by the Fourier sum of the hopping matrix 
elements $h_d$. We assume that the model 
has an energy gap $\Delta$ between the ground state and the minimum of the single-particle 
band. The general two-particle interaction is described in real space 
by the matrix elements $V_{d_1,d_2,d_3}$. 
Interactions among more particles, for example genuine three-particle interactions, are shown 
as dots in \eqref{eq.model_ham}  and can appear in the model, but are neglected in our approximation.

We assume that the Hamiltonian $H_0$ describes the 
hopping and interaction of conserved particles, i.e., the number of particles (hard-core bosons)
does not change. In general, a microscopic Hamiltonian will not have this property,
for instance if it is derived from a spin model \cite{sachd90}. But one can map such
microscopic non-conserving Hamilton operators to \textit{effective} Hamilton operators which
conserve the particle number.
There exists a variety of methods in literature which can be used to obtain such an 
effective Hamiltonian from a  general Hamiltonian 
\cite{solyo79,metzn12,wegne94,glaze93,glaze94,knett00a,knett03a,kehre06,fisch10,fause13, 
shast81a,uhrig99a,haege13,keim15,vande15}. 
Hence, here,  we do not consider this step, but rather discuss the general properties
of \eqref{eq.model_ham}.

\begin{figure}[ht]
\centering
\includegraphics[width=0.2\columnwidth]{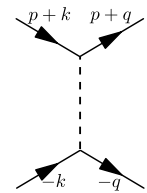}
\caption{Diagrammatic representation of the interaction vertex in
Eq.\ \eqref{eq.new_interaction4}}
\label{image:new_interaction4}
\end{figure}

Transforming the interaction into momentum space yields the interaction 
vertex including the on-site hard-core repulsion $U$
\begin{align} 
\label{eq.new_interaction4}
V(p,k,q) &= \frac{U}{N} \nonumber\\
 &+ \frac{1}{N} \sum\limits_{d_1,  d_2,  d_3} V_{d_1,d_2,d_3}  
e^{-i d_2 p} e^{-i d_1 k} e^{i d_3 q},
\end{align}
where $U$ is taken  later to infinity to implement the hard-core property.
The corresponding diagram is represented in Fig.\ \ref{image:new_interaction4}.
We stress that the additional interactions, i.e., all terms proportional to $V$,  in 
Eq.\ \eqref{eq.new_interaction4} will depend on the momenta $k$ and $q$.
Defining the momentum dependent vector
 \begin{align}
 \mathbf{f}^\dagger(k) = 
( 1 , e^{ik} , e^{-ik} , e^{i2k} , e^{-2ik} , \cdots),
 \end{align}
we can rewrite the interaction vertex as a bilinear form
\begin{align}
\label{eq.V_as_bilinear}
V(p,k,q) = \frac{1}{N} \mathbf{f}^\dagger(k) \underline{\underline{\Gamma}}_{0}(p) \mathbf{f}(q).
\end{align}
The advantage of this notation is that the dependencies on the momenta 
$p,k$, and $q$ are factorized.
The first few entries of the matrix  $\underline{\underline{\Gamma}}_{0}(p)$
read
\begin{subequations}
\begin{align}
 \underline{\underline{\Gamma}}_{0}(p) &=
 \\ \nonumber
   & \begin{pmatrix}
 U  & 0 & 0 & \cdots \\
 0 & \sum\limits_{d_2} V_{1,d_2,1} e^{-i d_2 p}  & 
\sum\limits_{d_2} V_{1,d_2,-1} e^{-i d_2 p} & \cdots \\
 0 & \sum\limits_{d_2} V_{-1,d_2,1} e^{-i d_2 p} & 
\sum\limits_{d_2} V_{-1,d_2,-1} e^{-i d_2 p} & \cdots \\
  \vdots & 					\vdots & 			\vdots & \ddots
 \end{pmatrix} \\
 &=:  \underbrace{\begin{pmatrix}
 U   & 0 & \cdots \\
 0   & 0 & \cdots \\
 \vdots & \vdots & \ddots
 \end{pmatrix}}_{\underline{ \underline{U}}} + \underline{\underline{V}} , 
\label{eq.blockmatrices}
\end{align}
\end{subequations}
where $\underline{ \underline{U}}$ and $\underline{ \underline{V}}$ 
are block matrices acting on different subspaces.

Below, we investigate the single-particle spectral 
function of the hard-core bosons defined by
\begin{align}
A(p, \omega) = \frac{-\mathrm{Im}}{\pi\sqrt{L}} 
\lim_{i \omega_\nu \rightarrow \omega + i 0^+} 
 \int_0^\beta \mathrm{d}\tau  e^{i \omega_\nu \tau} 
 \sum_j e^{-i p j} G(j, \tau) ,
\end{align}
where $G(j, \tau)$ is the single-particle temperature Green function 
\begin{align}
G(j, \tau) = - \left\langle T\left\{ b_{j}^\dagger(-i\tau) 
b_{0}^{\phantom\dagger}(0) \right\}\right\rangle .
\end{align}
and $L$ the system size. The Matsubara frequencies are denoted by 
$\omega_\nu = 2\nu \pi / \beta$ where $\beta = 1/T$ is the 
inverse temperature setting the Boltzmann constant to unity.
The spectral function is connected to the dynamic structure 
factor by means of the fluctuation-dissipation theorem
\begin{align}
\label{eq.fluctuation_dissipation}
S(p, \omega) = \frac{1}{1-e^{-\beta\omega}} \left[ A(p, \omega) + A(p, -\omega) \right],
\end{align}
which is directly accessible in inelastic neutron scattering experiments.

\section{Br\"uckner approach}
\label{sec.method}

In this section, we show how additional interactions can be included within the 
Br\"uckner approach. To keep the presentation transparent
we describe the procedure for a single flavor of hard-core bosons per site. In App.\ \ref{app.c} 
we show how the approach  can be extended to include multi-flavored bosons
 such as triplons in  dimerized spin systems.

The Br\"uckner approach is a low-temperature approximation for the 
single-particle spectral function based on diagrammatic perturbation theory. 
The key idea is to replace the hard-core bosonic creation and annihilation 
operators by normal bosonic operators and to enforce the hard-core constraint 
by an on-site repulsion $U$ which is taken to infinity in the end. The expansion 
parameter of the theory is the low density of excitations which is given 
proportional to $\exp(-\beta \Delta)$.  
  
\begin{figure}[ht]
\centering
\includegraphics[width=0.98\columnwidth]{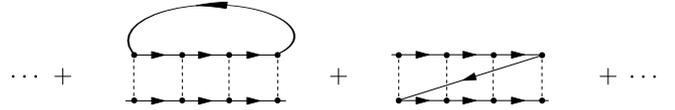}
\caption{Ladder diagrams with the interaction vertex given in 
Eq.\ \eqref{eq.new_interaction4}. }
\label{image:Ladder5_example}
\end{figure}

In leading order, all diagrams with a single propagator running backwards in 
imaginary time must be included. This leads to the summation of ladder diagrams 
as shown in Fig.\ \ref{image:Ladder5_example}. Here we extend the
previous approach \cite{fause14,fause15,strei15} by including also the 
additional interaction matrix elements $V_{d_1,d_2,d_3}$ in the diagrammatic 
ladders. In this way, we can explore how the additional interactions affect
the single-particle properties.

\begin{figure}[ht]
\centering
\includegraphics[width=0.9\columnwidth]{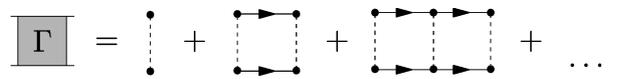}
\caption{Graphical definition of the scattering amplitude $\Gamma$. }
\label{image:Ladder5}
\end{figure}

\begin{figure}[ht]
\centering
\includegraphics[width=0.9\columnwidth]{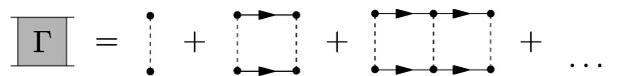}
\caption[Bethe-Salpeter equation including additional interactions]
{Bethe-Salpeter equation for the ladder diagrams in Fig.\ \ref{image:Ladder5}.}
\label{image:Bethe-Salpeter4}
\end{figure}

In a first step, we calculate the scattering amplitude $\Gamma$ as defined 
graphically in Fig.\ \ref{image:Ladder5}. The scattering amplitude 
describes the complete scattering of two particles.
It can be found as solution of the 
Bethe-Salpeter equation depicted in Fig. \ref{image:Bethe-Salpeter4} 
and denoted explicitly 
\begin{align} 
\label{eq.Bethe_Full_interaction}
\Gamma (P,K,Q) &= \frac{V(p,k,q)}{\beta} \\ \nonumber
&- \frac{1}{\beta} \sum\limits_L \Gamma(P,K,L) G(P+L) G(-L) V(p,l,q).
\end{align}
The capital letters are shorthand for the momentum and the Matsubara frequency, 
e.g., $P=(p,i\omega_p)$. We stress that due to the additional interactions
$\Gamma$  also depends on the relative momenta $K$ and $Q$, making the 
integration more challenging.

Since the dependence of the elementary interaction vertex $V(p,k,q)$ on the momenta
factorizes, we apply the same ansatz to the scattering amplitude
\begin{align}
\label{eq.Ansatz_Gamma}
\Gamma(P,K,Q) = \frac{1}{N} \mathbf{f}^\dagger(k) \uuline{\Gamma}(P) \mathbf{f}(q) 
\end{align}
which implicitly defines the scattering matrix $ \uuline{\Gamma}(P)$.
Inserting Eqs.\ \eqref{eq.V_as_bilinear} and \eqref{eq.Ansatz_Gamma} into 
Eq.\ \eqref{eq.Bethe_Full_interaction} and separating the dependence on 
the momenta yields
\begin{align}
\uuline{\Gamma}(P) &= \uuline{\Gamma}_0(p) 
\\ \nonumber
&-  \frac{1}{N \beta} \sum\limits_L 
\uuline{\Gamma}(P) G(P+L) G(-L) \mathbf{f}(l) \mathbf{f}^\dagger(l) \uuline{\Gamma}_0(p).
\end{align} 
This equation is the generalization of the scalar Bethe-Salpeter equation
for the scattering amplitude to the matrix-valued Bethe-Salpeter equation
for the scattering matrix  $\uuline{\Gamma}(P)$.
In addition, we also define the matrix
\begin{align}
\uuline{N}(P) := \frac{1}{N} \sum\limits_l \mathbf{f}(l) \mathbf{f}^\dagger(l) M(P,l),
\end{align}
where we used the scalar function
\begin{align}
M(P,l) &:= \frac{1}{\beta} \sum\limits_{i\omega_l} G(P+L) G(-L).
\end{align}

Since the frequency dependence of $M(P,l)$ is in $\mathcal{O}(\omega_{p}^{-1})$, the 
matrix $\uuline{N}(P)$ has a spectral Hilbert representation.
We denote its spectral function by $\uuline{\rho}(p, \omega)$.  
Inserting the previous definitions into the Bethe-Salpeter equation 
\eqref{eq.Bethe_Full_interaction} yields the matrix expression
\begin{align}
\uuline{\Gamma}(P) = \uuline{\Gamma}_0(p) - \uuline{\Gamma}(P) \uuline{N}(P) 
\uuline{\Gamma}_0(p),
\end{align}
which represents a geometric series for matrices.
It can be easily solved by the scattering matrix
\begin{align}
\label{eq.Matrix_inversions}
\uuline{\Gamma}(P) &= \left(  \uuline{\Gamma}_{0}^{-1}(p) + \uuline{N}(P)   
\right)^{-1}.
\end{align}
In analogy to the scalar case \cite{fause14} the
spectral representation of $\uuline{\Gamma}(P)$ has two contributions: 
(i) A high-energy contribution stemming from a virtual antibound
state at $\omega \approx U$ and
(ii) a low-energy contribution, where $\omega \approx \Delta$. 
In the following two subsections we will determine the exact form of these 
contributions taking the additional interaction into account.

\subsection{High-energy contribution}

The aim is to perform the limit $U\to\infty$ analytically. At low energies, one
can set $U=\infty$ in the equations and evaluate them straightforwardly. 
But there also arise contributions from an antibound state at high energies.
The analytical determination of these contributions required some care
for the on-site repulsion. Due to the additional interaction, these contributions are
modified as we explain now. 

To compute the exact contribution we need to calculate the 
position of the pole in the spectral representation of $\uuline{\Gamma}(P)$ 
for $\omega \approx U$. The pole can be obtained from the zero eigen value
of the matrix inverse of Eq.\ \eqref{eq.Matrix_inversions}
\begin{align} \label{eq.denominator}
 \uuline{\Gamma}(P)^{-1}  = \uuline{\Gamma}_{0}^{-1}(p) + \uuline{N}(P) .
\end{align}
Note, that 
\begin{align}
\uuline{\Gamma}_{0}^{-1}(p) = \uuline{U}^{-1} + \uuline{V}^{-1}
\end{align}
holds where it is understood that 
the matrix inverses of $\uuline{U}$ and $\uuline{V}$ are
in the respective subblocks where the matrices contribute, see Eq.\ \eqref{eq.blockmatrices}.
This means that $\uuline{U}^{-1}$ only has an entry in the $(1,1)$ matrix element, and
$\uuline{V}^{-1}$ only for matrix elements $(m,n)$ with $m,n>1$.

If we expand $\uuline{N}(P)$ for $\omega \rightarrow \infty$ we obtain
\begin{align}
\uuline{N}(P) = \frac{\uuline{\rho}_0(p)}{\omega} + \frac{\uuline{\rho}_1(p)}{\omega^2} 
+ \mathcal{O} \left( \frac{1}{\omega^3} \right),
\end{align}
where $\uuline{\rho}_m(p)$ denotes the $m^\mathrm{th}$ moment in $x$ of the 
matrix-valued spectral function $\uuline{\rho}(p,x)$. This is in complete
analogy to the scalar case in Refs.\ \cite{fause14,fause15}, but generalized here
to matrices.

We introduce the parametrization $\omega = \bar{\omega} U$, such that 
$\bar{\omega} = \mathcal{O}(1)$ for $U\to \infty$.
Inserting the expansion of $\uuline{N}(P)$ into Eq.\ \eqref{eq.denominator} yields
\begin{equation}
\begin{aligned}
\uuline{U}^{-1} + \uuline{V}^{-1} + \frac{1}{U\bar{\omega}} \uuline{\rho}_0(p) 
+ \frac{1}{U^2 \bar{\omega}^2} \uuline{\rho}_1(p) \\
= \uuline{V}^{-1}  + \frac{1}{U} \underbrace{\left( \frac{1}{\bar{\omega}}  
\uuline{\rho}_0(p) + U \uuline{U}^{-1} \right)}_{X_1} + \frac{1}{U^2} 
\underbrace{\left( \frac{\uuline{\rho}_1(p)}{\bar{\omega}^2 } \right)}_{X_2} .
\end{aligned}
\end{equation}
To determine the correct contribution from the antibound state at frequencies 
$\omega \approx U$, we need to calculate the scattering matrix in 
this frequency range.  For this purpose, we employ matrix perturbation theory in 
the parameter $1/U$. The zeroth order is given by $\uuline{V}^{-1}$. The 
perturbations are the matrices $X_1$ and $X_2$ in first and second order, respectively.
We denote by $(\lambda_i, \mathbf{e}_i)$ the unperturbed eigen pair, i.e.,
eigen value and corresponding eigen vector, of $\uuline{V}^{-1}$. 
The first eigen pair represents the important contribution of the
antibound state and reads
\begin{align}
\label{eq.eigenpair}
\lambda_1 = 0 \, , \quad \mathbf{e}_1 = (1,0,0,\dots)^\mathrm{T}.
\end{align}
But for any finite $U$,  higher order contributions mix into this eigen pair. 
We refer the reader to App.\ \ref{app.b} for the discussion of the matrix 
perturbation theory which is essentially standard first and second order
perturbation theory from any quantum mechanics text book. 

\begin{figure}[ht]
\vspace*{10pt}
\centering
\includegraphics[width=0.65\columnwidth]{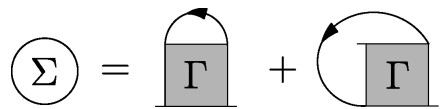}
\caption{Self-energy diagrams obtained from summing 
the scattering matrix $\Gamma(P,K,Q)$ with another dressed propagator.}
\label{image:Self-energy}
\end{figure}

Once the scattering matrix has been calculated, 
the self-energy contribution from the antibound state at high energy (denoted `he') 
can be  calculated by closing the scattering matrix by another propagator, see
arrowed propagator in Fig.\ \ref{image:Self-energy}. The first diagram on the right hand side
in  Fig.\ \ref{image:Self-energy} corresponds to the Hartree contribution and 
the second to the Fock contribution. Explicitly, they are given by
\begin{align}
&\Sigma_{\mathrm{he}}(P) =\\ \nonumber
& \sum\limits_{K} G(K) \left[ \Gamma(P+K,-P,-P) + \Gamma(P+K,-K,-P) \right] .
\end{align}
The explicit sum over the  Matsubara frequencies is similar to the one 
in the scalar case \cite{fause14,fause15}. 
Inserting the expression for the scattering amplitude yields for the Hartree contribution
\begin{widetext}
\begin{align}
\Sigma_{\mathrm{Hart, he}}(P) =
-\frac{1}{N}  \sum\limits_{k} \int\limits_{-\infty}^{\infty}   \mathrm{d}x'
		 \frac{A_k(x')U^2 \rho^{11}_{0}(p+k) \left[ 1- \frac{1}{U} 
		\sum\limits_{j \neq 1} \frac{1}{\rho_0^{11}(p+k)} 2 \mathrm{Re} 
		\left( W_j(p+k)^* f_j(-p) \right) \right]}{i \omega_p - 
		(\omega_U(p+k)- x')} \frac{1}{e^{\beta x'}-1} .
\end{align}
where the functions $W_j$ (and $V_j$ for later use), 
$f_j$ and $\omega_U$ are defined in App.\ \ref{app.b}.
The Hartree and Fock contribution of the pure hard-core repulsion reads
\begin{align}
\Sigma_{\mathrm{Hart}, U} = \Sigma_{\mathrm{Fock}, U} = \frac{1}{N} \sum\limits_{k} \int\limits_{-\infty}^{\infty} \mathrm{d}x' A_k(x') U \frac{1}{e^{\beta x'}-1} .
\end{align}
Expanding the high energy Hartree term in $1/U$ and combining it with $\Sigma_{\mathrm{Hart}, U}$
of the pure hard-core repulsion we eventually perform the limit $U\to\infty$  and obtain
\begin{equation}
\begin{aligned}
\Sigma_{\mathrm{Hart, he}}(P) + \Sigma_{\mathrm{Hart}, U} = -\frac{1}{N}  
\sum\limits_{k} \int\limits_{-\infty}^{\infty}  \mathrm{d}x' A_k(x') 
\left[\frac{-i \omega_p}{\rho^{11}_{0}(p+k)} - \frac{\sum\limits_{i \neq 1} V_i(p+k)}
{ {\rho^{11}_{0}}^2(p+k)} \right. 
\\
 \left. - \frac{\rho^{11}_{1}(p+k)}{{\rho_0^{11}}^2(p+k)}+\frac{x'}{\rho^{11}_{0}(p+k)} 
-\sum\limits_{j \neq 1} \frac{1}{\rho_0^{11}(p+k)}  2 \mathrm{Re} \left( W_j(p+k)^* f_j(-p)
 \right) \right] \frac{1}{e^{\beta x'}-1}. 
\end{aligned}
\end{equation}
Note that this results
directly reflects the relevant expression for the hard-core particles.

The Fock term can be computed in a similar fashion yielding
\begin{equation}
\begin{aligned}
 &\Sigma_{\mathrm{Fock, he}}(P) + \Sigma_{\mathrm{Fock}, U} = 
-\frac{1}{N}  \sum\limits_{k} \int\limits_{-\infty}^{\infty}  
\mathrm{d}x' \frac{A_k(x')}{e^{\beta x'}-1}  \left[\frac{-i \omega_p}{\rho^{11}_{0}(p+k)} 
- \frac{\sum\limits_{i \neq 1} V_i(p+k)}{ {\rho^{11}_{0}}^2(p+k)} \right. 
\\ 
& \left. 
- \frac{\rho^{11}_{1}(p+k)}{{\rho_0^{11}}^2(p+k)}+\frac{x'}{\rho^{11}_{0}(p+k)}  
-\sum\limits_{j \neq 1} \frac{1}{\rho_0^{11}(p+k)} \left( W_j(p+k)^* f_j(-p) 
 +W_j(p+k) f_j(-k)^* \right) \right] .
\end{aligned}
\end{equation}
\end{widetext}
Comparing these expressions to the ones in the case of a pure hard-core repulsion, see
Eq.\ (A7) in Ref.\ \onlinecite{fause14}, 
we  see that two additional contributions to the real part of the self-energy 
arise from the additional interactions.

\subsection{Low-energy contribution}

In the low-energy sector, we can take the limit $U\rightarrow \infty$ directly
without considering intricate limits. 
Then, the matrix $\uuline{\Gamma_0}^{-1}$ equals $\uuline{V}^{-1}$.
We use the Hilbert representation of $\uuline{N}$ to calculate the Hilbert representation 
of $\uuline{\Gamma}(P) - \uuline{V}(p)$, i.e., the ladder diagrams minus the 
simple Hartree-Fock diagrams which are constant in frequency
\begin{subequations}
\begin{align}
\uuline{\Gamma}(P) - \uuline{V}(p) &= \int\limits_{-\infty}^\infty \mathrm{d}x' 
\frac{\uuline{\rho}_\Gamma(p, x')}{i \omega_p -x'}  ,\\
\uuline{\rho}_\Gamma(p, \omega) &= \begin{pmatrix}
			\rho^{11}_{\Gamma} & \rho^{12}_{\Gamma} & \cdots \\
			\rho^{21}_{\Gamma} & \rho^{22}_{\Gamma} & \cdots \\
			\vdots & \vdots & \ddots
			\end{pmatrix} \\
			&= \frac{-\mathrm{Im}}{\pi}  \lim\limits_{i \omega_p \rightarrow \omega} 
			\left( \uuline{V}^{-1} + \uuline{N}(P) \right)^{-1} - \uuline{V}.
\end{align}
\end{subequations}
In general, these expressions cannot be simplified further
analytically. For a given interaction, however,
the spectral representation can be obtained numerically for fixed
frequency and momentum.

Next, we determine the low-energy (le) contributions to the self-energy. Similar to 
the high-energy contributions, there are Hartree- and the Fock-like contributions
which result from the two ways to close the scattering amplitude, see Fig.\
\ref{image:Self-energy}.

\begin{widetext}
They appear in the two terms in
\begin{subequations}
\begin{align}
\Sigma_\mathrm{le}(P) &= \sum\limits_{K} G(K) ( \Gamma(P+K,-P,-P) + \Gamma(P+K,-K,-P) ) 
\\
&= \frac{-1}{N \beta} \sum\limits_{k, i \omega_k} \int\limits_{-\infty}^\infty 
\int\limits_{-\infty}^\infty  \mathrm{d}x' \mathrm{d}x'' \frac{A_k(x')}{i \omega_k - x'}
\left( \mathbf{f}^\dagger(-p) + \mathbf{f}^\dagger(-k)\right) 
\frac{\uuline{\rho}_\Gamma(p+k, x'')}{i \omega_p + i \omega_k -x''}  \mathbf{f}\left(-p\right) ,
\end{align}
\end{subequations}
The first term represents the Hartree contribution and the second the Fock contribution
of the scattering amplitude.
We substitute $k \rightarrow k-p$ and sum over all Matsubara frequencies
in order to obtain
\begin{align}
\Sigma_\mathrm{le}(P) &= \frac{1}{N} \sum\limits_k \int\limits_{-\infty}^\infty 
\mathrm{d}x'' \left[ \mathbf{f}^\dagger\left(-p\right) + 
\mathbf{f}^\dagger\left(-(k-p)\right) \right] A_{k-p}(x''-\omega) 
\uuline{\rho}_\Gamma(k, x'')  \mathbf{f}\left(-p\right)  
 \left[ \frac{1}{e^{\beta (x''-\omega)} - 1} - \frac{1}{e^{\beta x''} - 1 } \right] .
\end{align}
\end{widetext}

\subsection{Hartree-Fock contributions for additional interactions}

We stress that the simple  Hartree- and Fock-contributions without any frequency dependence
cannot be expressed by spectral representations. Hence they must be dealt with separately.
For the additional interaction the simple Hartree-term reads
\begin{align}
\Sigma_{\mathrm{Hart}, V} &= - \sum\limits_{k} \int\limits_{-\infty}^{\infty}  
\mathrm{d}x'  \frac{A_k(x')}{e^{\beta x'}-1} V(p+k,-p,-p) .
\end{align}
The Fock-term for the additional interaction is given by
\begin{align}
\Sigma_{\mathrm{Fock}, V} &= - \sum\limits_{k} \int\limits_{-\infty}^{\infty}  
\mathrm{d}x' \frac{A_k(x')}{e^{\beta x'}-1} V(p+k,-k,-p) .
\end{align}
In the multi-flavor case, the Hartree contribution is multiplied by the number of flavors $N_\mathrm{f}$.

\subsection{Self-energy and spectral function}

Now we are in the position to sum all contributions to the self-energy in 
leading order in $\exp(-\beta \Delta)$
\begin{equation}
\begin{aligned}
\Sigma &= \Sigma_\mathrm{le} + \Sigma_{\mathrm{Fock}, V} + 
\Sigma_{\mathrm{Hart}, V} + \Sigma_{\mathrm{Fock, he}} 
\\ 
&+ \Sigma_{\mathrm{Fock}, U} + \Sigma_{\mathrm{Hart, he}} 
+ \Sigma_{\mathrm{Hart}, U}.
\end{aligned}
\end{equation}
We omitted the dependence on the total momentum and frequency $P$ for the sake of brevity. 
The only terms which are not affected by the additional interaction are the simple
Hartree- and Fock-Terms $\Sigma_{\mathrm{Hart}, U} + \Sigma_{\mathrm{Fock}, U}$ of the pure 
hard-core repulsion.
All other contributions include terms which are 
proportional to $V_{d_1,d_2,d_3}$.

Once the self-energy is calculated we can determine the spectral function 
using the Dyson equation
\begin{align}
\label{eq:dyson}
&  A(p, \omega) = \nonumber\\
& \quad \frac{-1}{\pi} \frac{\mathrm{Im} \Sigma(\omega,p)}
{\left(\omega - \omega(p) - \mathrm{Re} \Sigma(\omega,p) \right)^2 
+ \left( \mathrm{Im} \Sigma(\omega, p) \right)^2} .
\end{align}
We point out that we sum the diagrams self-consistently, i.e., all propagators
are dressed propagators. In practice, we start from an 
initial guess for the propagators, compute the self-energy, 
insert it in the Dyson equation \eqref{eq:dyson} to determine the
propagators. This cycle is iterated as long as the ensuing
propagators differ sizeably from the input propagators .
As a numerical criteria we use the first and second moment of the spectral function. Once they do not
change anymore within machine precision the iteration is stopped and the result is considered to be converged for practical purposes.

\section{Results for Heisenberg spin ladders}
\label{sec.results}

In this section, we apply the developed Br\"uckner approach including
additional interactions to a generic system, namely the 
dimerized Heisenberg spin $S=1/2$ ladder. It is well established that the
elementary excitations in this system are hard-core bosons with three
flavors of triplet character \cite{knett01b} called triplons \cite{schmi03c}.
At zero temperature, much is known
about the system \cite{schmi05b} and the agreement between experiment
and theory is quantitative \cite{windt01,notbo07}. Hence, it is 
confirmed that the effective Hamiltonian describing the motion and the
interaction of the elementary triplons is known, in particular
the additional NN and NNN interactions \cite{krull12}.

In order to illustrate the applicability and usefulness of the
extended Br\"uckner approach, we analyze the influence of these 
additional interactions on the spectral functions of the
spin ladder at finite temperatures. 
This offers the opportunity to explore the feedback effect of strong correlations 
in the two-particle sector on the single-particle mode at finite temperature.

\subsection{Model of the Heisenberg spin ladder}

\begin{figure}[ht]
\centering
\includegraphics[width=0.9\columnwidth]{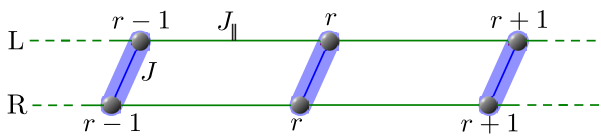}
\caption[Exchange couplings in the Heisenberg ladder]{Exchange couplings in the dimerized Heisenberg ladder.}
\label{image:Spin_Ladder2}
\end{figure}

The Hamiltonian of the Heisenberg spin ladder is illustrated in 
Fig.\ \ref{image:Spin_Ladder2}. Its explicit form expressed in 
spin operators  reads
\begin{align}
H = \sum\limits_{r} J \mathbf{S}_{r}^{\mathrm{L}} \cdot 
\mathbf{S}_{r}^{\mathrm{R}} + J_{\parallel} \left( \mathbf{S}_{r}^{\mathrm{R}} 
\cdot \mathbf{S}_{r+1}^{\mathrm{R}} + \mathbf{S}_{r}^{\mathrm{L}} 
\cdot \mathbf{S}_{r+1}^{\mathrm{L}} \right),
\end{align}
where $J$ is the coupling on the strong bonds on the rungs defining the dimers and 
$x = J_{\parallel}/J$  is the relative strength of the interdimer coupling
along the legs of the ladder.
The index $r$ denotes the dimer sites, and L and R refer to the left and right leg 
of the ladder respectively.

It was shown in Ref.\ \cite{krull12} that deepCUT provides an excellent 
renormalization tool to compute the effective model in terms of triplons 
for gapped dimerized spin systems which conserve the triplon number.  
Thus we use a deepCUT calculation to 
obtain the hopping and interaction matrix elements of the triplons. 
The resulting effective Hamiltonian in terms of triplon creation and annihilation
operators reads
\begin{align}
\frac{H_\mathrm{eff}}{J} &= E_0 + H_1 + H_2
\end{align}
where $E_0$ is the ground state energy, $H_1$ the one-triplon Hamiltonian,
and $H_2$ the two-triplon interaction. The one-triplon Hamiltonian describes
the motion of triplons via hopping processes over distance $d$
\begin{align}
\label{eq:H1}
H_1 &= \sum_{r, \alpha} h_0 t_{r,\alpha}^\dagger t_{r,\alpha}^{\phantom\dagger} + \sum_{r,|d| > 0} 
\sum_\alpha \frac{h_d}{2} t_{r+d,\alpha}^\dagger t_{r,\alpha}^{\phantom\dagger},
\end{align}
where $\alpha\in \left\lbrace x,y,z\right\rbrace$ is the flavor index and 
$t_{r,\alpha}^\dagger, t_{r,\alpha}$ are triplon creation and annihilation operators,
respectively. Note that we let all triplons hop in the same way 
due to spin rotation invariance
and inversion symmetry fixes $h_d=h_{-d}$ being real. The ensuing dispersion
$\omega(k)$ reads 
\begin{align}
 \omega(k)&= \sum_{d \geq 0} h_d \cos(dk),
\end{align}
where we set the lattice constant to unity. In this paper, we restrict the
hopping range to $|d|\le 6$ for simplicity. This is completely sufficient
to describe the dispersion of spin ladders to good accuracy up to $x=1$.

The two-triplon Hamiltonian describes the additional interactions
which are the focus of our work. We restrict them to the processes which
can arise up to order 2 in $x$, see Tab.\ III in Ref.\ \onlinecite{krull12}
\begin{subequations}
\label{eq:H2}
\begin{align}
H_2 &=  w_1 \sum_{r} \sum_\alpha
t_{r,\alpha}^\dagger t_{r+1,\alpha}^\dagger 
t_{r+1,\alpha}^{\phantom\dagger} t_{r,\alpha}^{\phantom\dagger} \\
& + w_2 \sum_{r} \sum_\alpha
t_{r+2,\alpha}^\dagger t_{r+1,\alpha}^\dagger 
t_{r+1,\alpha}^{\phantom\dagger} t_{r,\alpha}^{\phantom\dagger} +\mathrm{h.c.} \\
&+ w_3 \sum_{r} \sum_{\alpha\ne\gamma}
t_{r,\alpha}^\dagger t_{r+1,\alpha}^\dagger 
t_{r+1,\gamma}^{\phantom\dagger} t_{r,\gamma}^{\phantom\dagger} \\
&+ w_4 \sum_{r} \sum_{\alpha\ne\gamma}
t_{r,\alpha}^\dagger t_{r+1,\gamma}^\dagger 
t_{r+1,\alpha}^{\phantom\dagger} t_{r,\gamma}^{\phantom\dagger} \\
&+ w_5\sum_{r} \sum_{\alpha\ne\gamma}
t_{r+2,\gamma}^\dagger t_{r+1,\alpha}^\dagger 
t_{r+1,\alpha}^{\phantom\dagger} t_{r,\gamma}^{\phantom\dagger}  +\mathrm{h.c.}\\
&+ w_6\sum_{r} \sum_{\alpha\ne\gamma}
t_{r,\gamma}^\dagger t_{r+1,\alpha}^\dagger 
t_{r+1,\alpha}^{\phantom\dagger} t_{r,\gamma}^{\phantom\dagger} \\
&+ w_7\sum_{r} \sum_{\alpha\ne\gamma}
t_{r+2,\gamma}^\dagger t_{r+1,\gamma}^\dagger 
t_{r+1,\alpha}^{\phantom\dagger} t_{r,\alpha}^{\phantom\dagger}  +\mathrm{h.c.}
\end{align}
\end{subequations}
We stress that the numerical prefactors $w_j$
are determined by a deepCUT calculation of order 6.
The precise numbers used are listed in Tab.\ \ref{tab:prefactors} for the values
of $x$ considered in this article.

\onecolumngrid

\begin{table}[htb]
\begin{tabular}{|c||c|c|c|c|c|c|c|}
\hline 
 $x$         & $h_0$ & $h_1$ & $h_2$ & $h_3$ & $h_4$ & $h_5$ & $h_6$  \\
\hline\hline
$0.2$ & $1.032449982$ & $0.197447202$ & $-0.011976902$ & $0.001148659$ & $-0.000146837$ & $0.000021155$ & $-0.000003167$\\ \hline
$0.6$ & $1.279046995$ & $0.508412978$ & $-0.131238627$ & $0.025917844$ & $-0.009368954$ & $0.003760173$ & $-0.001417178$\\ \hline
$0.8$ & $1.447703192$ & $0.591395735$ & $-0.235167175$ & $0.048815082$ & $-0.021795289$ & $0.010694648$ & $-0.004713291$\\ \hline
$1.0$ & $1.627729418$ & $0.643483774$ & $-0.352625277$ & $0.075238493$ & $-0.037946270$ & $0.021098310$ & $-0.010197167$\\ \hline
\end{tabular}

\medskip

\begin{tabular}{|c||c|c|c|c|c|c|c|}
\hline 
 $x$         & $w_1$ & $w_2$ & $w_3$ & $w_4$ & $w_5$ & $w_6$ & $w_7$  \\
\hline\hline
$0.2$ & $-0.010542152$ & $0.011321380$ & $-0.094559805$ & $0.099997195$ & $0.005991517$ & $-0.015979524$ & $0.005378499$\\ \hline
$0.6$ & $-0.063956523$ & $0.103995401$ & $-0.249703472$ & $0.299185869$ & $0.066627091$ & $-0.113405271$ & $0.044179934$\\ \hline
$0.8$ & $-0.071899966$ & $0.168824957$ & $-0.315550868$ & $0.397077500$ & $0.120736262$ & $-0.153387044$ & $0.068217261$\\ \hline
$1.0$ & $-0.066366893$ & $0.235731322$ & $-0.376683724$ & $0.492295270$ & $0.182036845$ & $-0.182288001$ & $0.092018083	$\\ \hline
\end{tabular}
\caption{
\label{tab:prefactors}
Upper table: Numerical values of the hopping matrix elements $h_d$, see
Eq.\ \eqref{eq:H1},
for various coupling ratios $x=J_\parallel/J$ determined by deepCUT.
Lower table: Numerical values of the interaction matrix elements $w_j$,
see Eq.\ \eqref{eq:H2},
for various coupling ratios $x=J_\parallel/J$ determined by deepCUT.
}
\end{table}

\twocolumngrid

To include the additional interactions among multi-flavored hard-core bosons, 
we must deal with two types of interaction vertices: (i) ingoing triplons with flavor 
$\gamma$ and outgoing triplons with flavor $\alpha$ which may or may not
be equal to $\alpha$. The value of the interaction vertex depends on 
$\alpha \neq \gamma$ or $\alpha =\gamma$. (ii) Two triplons with flavor $\alpha \neq \gamma$ go in and come out, i.e., they interact with each other. 
In the first case, Hartree- and Fock-like diagrams contribute
to the self-energy. In the second case, only Hartree-like diagrams contribute. Their
contribution acquires a prefactor of $2$ due to the fact that there 
are two possible flavors for the Green function in the closed loop.
Hence, we need two types of interaction matrices:  $ \uuline{V}^{\alpha \alpha \gamma \gamma}$ 
for the type (i) interactions and $ \uuline{V}^{\alpha \gamma \alpha \gamma}$ 
for the type (ii) interactions.

In our approach we include additional interaction of the form
occurring in second order in $x$. This determines the type of quartic terms shown
in \eqref{eq:H2}; their prefactors are determined by deepCUT
so that higher order contributions are included as well. The terms occurring
in \eqref{eq:H2} have at maximum a spatial range of 2, i.e., besides the 
rung $r$ the farthest rung addressed is $r\pm 2$. This implies that the matrices 
$\uuline{\Gamma}_0(p)$ and $\uuline{\Gamma}(P)$ are finite and can be treated 
numerically.

Among the ladder diagrams for the self-energy we 
can distinguish three different types: 
(a) Fock-like diagrams with interaction matrix 
$\uline{\uline{U}} + \uuline{V}^{\alpha \alpha \gamma \gamma}$, (b) Hartree-like 
diagrams with interaction matrix 
$\uline{\uline{U}} + \uuline{V}^{\alpha \alpha \gamma \gamma}$, and
(c) Hartree-like diagrams with the interaction matrix $\uline{\uline{U}} + 
\uuline{V}^{\alpha \gamma \alpha \gamma}$. In the latter case, 
$\alpha\neq\gamma$ is implied.
The diagrams (a) and (b) can be treated in the same way in 
the Bethe-Salpter equation, but yield different contributions 
on the level of the self-energy due to the different final
sum over  the last propagator (arrow in Fig.\ \ref{image:Self-energy}).  
  
\subsection{Results}

\begin{figure}[htb]
\centering
\includegraphics[width=1.0\columnwidth]{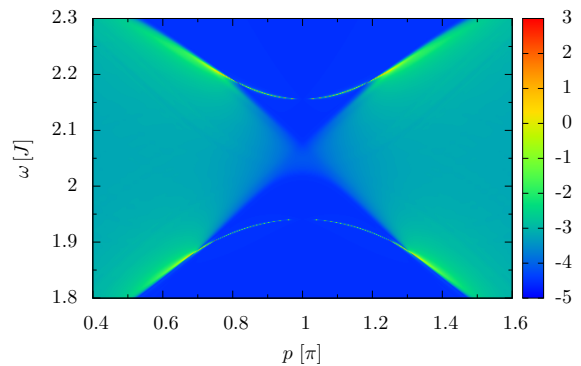}
\caption[Scattering matrix element]{Diagonal (2,2) matrix element of the spectral 
function $\uuline{\rho}_\Gamma^{x y x y}$ of the scattering matrix
for $x=0.2$ at $T=0.3J$ as function of total momentum $p$ and frequency $\omega$. 
The color axis has a logarithmic scale to depict both strong peaks and weak continua. 
The arcs below and above the continua around $p=\pi$ stem from the
triplet bound state and the quintuplet antibound state, respectively.
The singlet bound state does not appear in this spectral response because it does
not have overlap with the $xy$ triplon pair.}
\label{image:Bound_States}
\end{figure}

The most striking effect of the additional interactions is the occurrence of 
bound and antibound states in the low-energy sector 
\cite{uhrig96b,uhrig96be,trebs00,knett01b,zheng01a}. These states appear because
the additional interactions imply either an attractive or repulsive net effect 
depending on the total spin $S_\mathrm{tot}$ of the pair of triplons under study.
For $S_\mathrm{tot}=0$, rather strong attraction is at work, for 
$S_\mathrm{tot}=1$ it is weaker by about a factor 2 and for $S_\mathrm{tot}=2$
the triplons repel each other. Effects of this binding and antibinding 
can be observed in the matrix elements of $\uuline{\Gamma}$. 
In Fig.\ \ref{image:Bound_States} we show a matrix element 
of the spectral function $\uuline{\rho}_\Gamma^{x y x y}$.

The spectral function is dominated by a two-particle continuum with a bound 
state below the continuum and an antibound state above the continuum at 
$p \approx \pi$.  The bound and antibound states coincide with $S_\mathrm{tot}=1$
and  $S_\mathrm{tot}=2$ excitations in the triplon language 
\cite{zheng01a, knett03b}. Note, 
that the $S_\mathrm{tot}=0$ bound state does not show up
because it has no overlap with 
the interaction matrix $\uuline{V}^{\alpha \gamma \alpha \gamma}$ for
$\alpha\neq\gamma$.  

Next, we investigate the spectral function $A(p,\omega)$ proportional to
the scattering rate of inelastic neutron scattering for finite temperatures 
and various values of the relative coupling strength $x$.
Our focus lies on an exemplary comparison of three kinds of results.
The first kind is the calculation for a pure hard-core bosonic system, 
i.e., only the infinite on-site repulsion is taken into account.
Its curves are denoted by `hard-core' in the following figures.
The second kind is a calculation including the additional interactions
on the level of a static Hartree-Fock mean-field calculation
as done previously for spin systems \cite{strei15,klyus16,fause16}.
Its curves are denoted by `mean field' in the following figures.
The third kind is the full calculation of the ladder diagrams considering
all interaction vertices including those of the additional interactions.
Its curves are denoted by `full Br\"uckner' in the following figures.

\begin{figure}[htb]
\centering
\includegraphics[width=1.0\columnwidth]{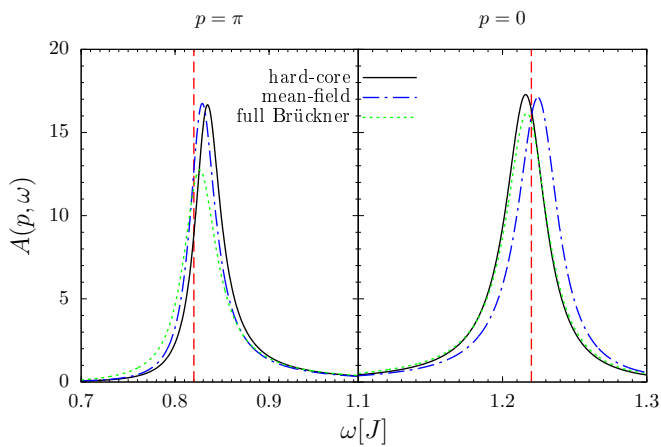}
\caption[Spectral functions  for $x=0.2$]{Spectral functions of 
for $x=0.2$ at momentum $p=\pi$ (left panels, gap mode) and $p=0$ 
(right panels, maximum mode). The temperature is  $T=0.3J$. 
The vertical dashed lines show where the $T=0$ $\delta$-peaks of the 
hard-core bosons are located. }
\label{fig:x02}
\end{figure}

Fig.\ \ref{fig:x02} starts the analysis by displaying
the spectral functions for $x=0.2$, i.e., the model for
which Fig.\ \ref{image:Bound_States} depicts the signatures
of (anti)bound states in the spectral functions of the scattering
matrix. Note that the spin ladder has its gap mode, i.e., the mode
with the lowest energy at momentum $p=\pi$ while its maximum mode,
i.e., the mode with maximum energy,
occurs at $p=0$ except for large values of $x\gtrapprox0.8$.
The left panels display the gap modes while the right panels
display  the mode at $p=0$. The differences between the
three kinds of calculations are still fairly small at $x=0.2$
as one might have expected due to the smallness of the corrections.
In particular, the broadening is clearly dominated by the
scattering due to the hard-core repulsion.
It must be noted, however, that the size of the interaction
\emph{relative} to the band width does not vanish for $x\to0$,
but stays finite. Interestingly, even the qualitative position
of the peak relative to the $T=0$ dispersion depends on 
the kind of calculation. The maximum mode (right panel in 
Fig.\ \ref{fig:x02}) in the hard-core calculation lies below the
zero-temperature energy, but above it in the mean-field  calculation while
the full Br\"uckner calculation brings it back to the hard-core calculation.

\begin{figure}[htb]
\centering
\includegraphics[width=1.0\columnwidth]{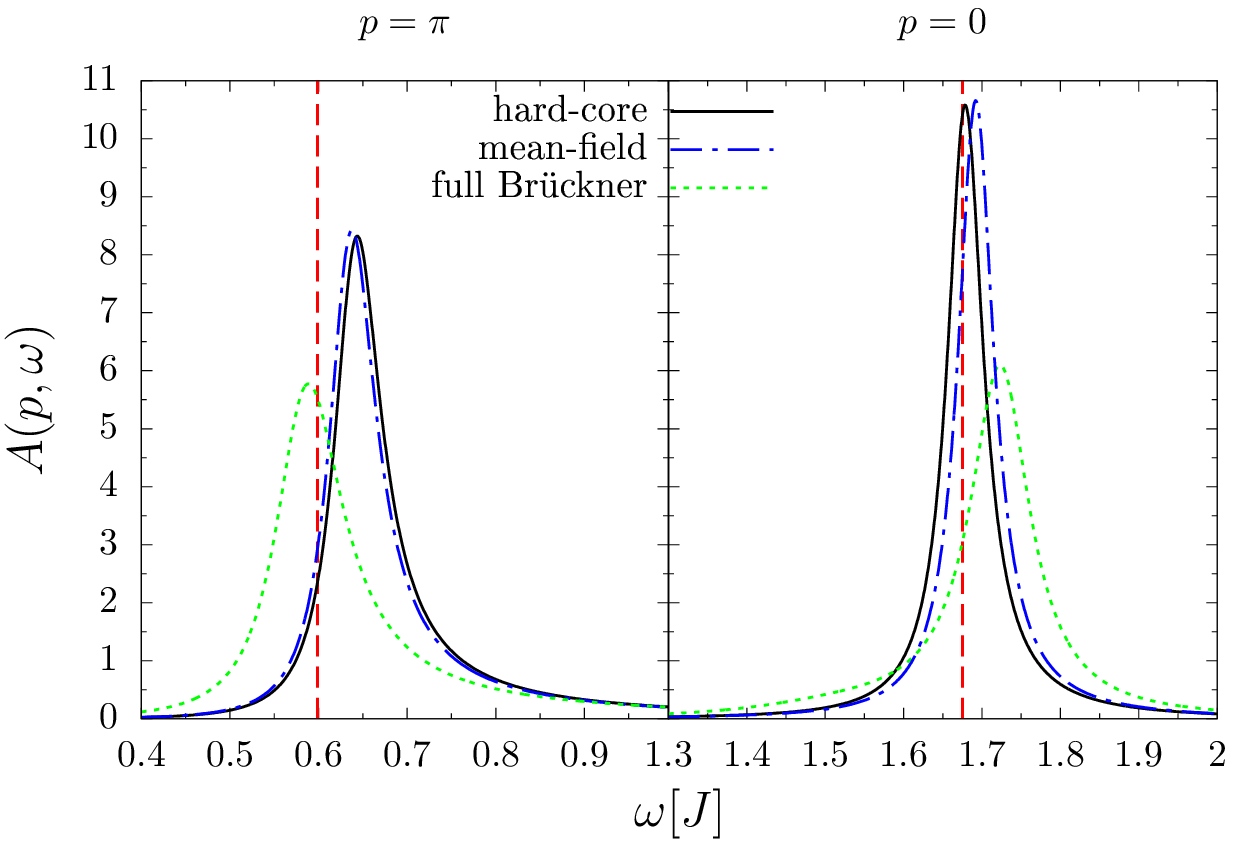}
\includegraphics[width=1.0\columnwidth]{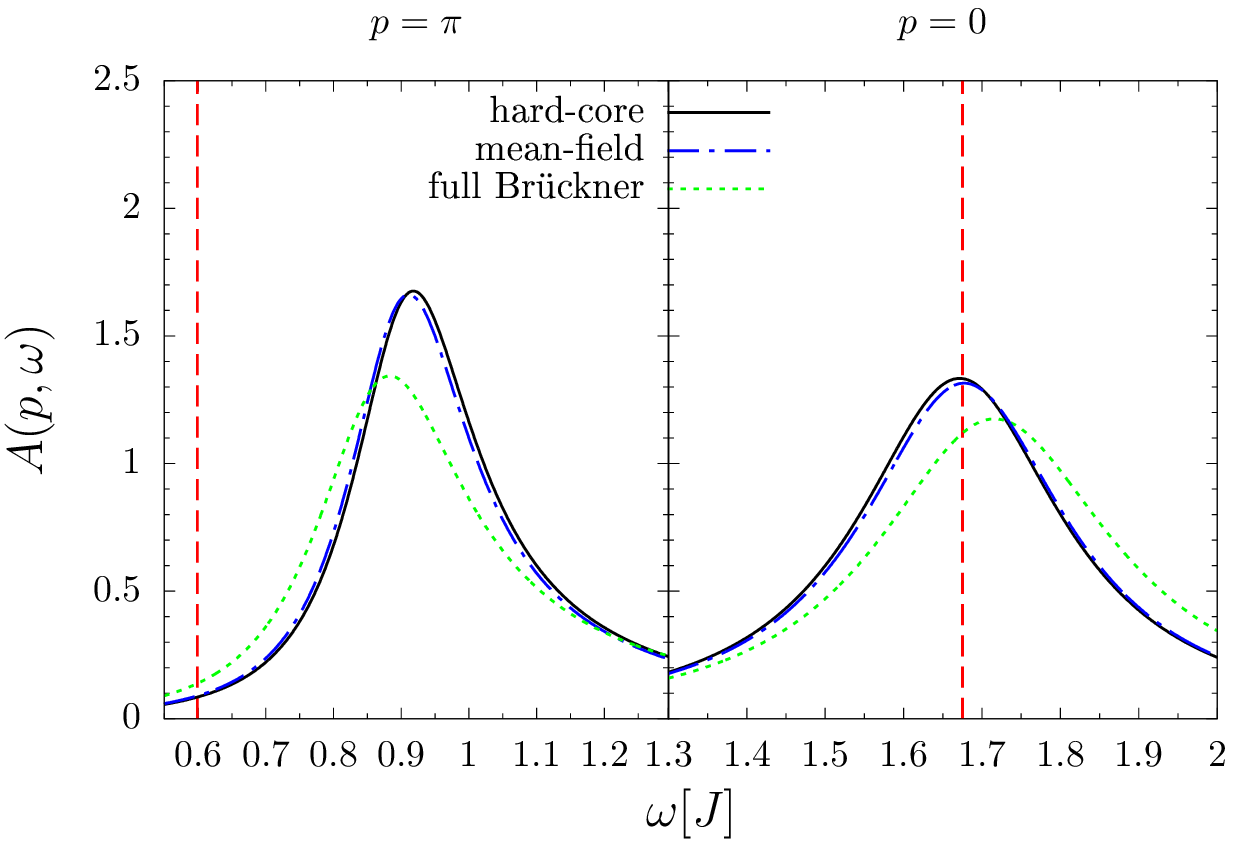}
\caption[Spectral functions  for $x=0.6$]{Spectral functions of 
for $x=0.6$ at momentum $p=\pi$ (left panels, gap mode) and $p=0$ 
(right panels, maximum mode). The temperature is  $T=0.3J$ for the
upper panels and $T=0.6J$ for the lower ones. The vertical dashed lines
show where the $T=0$ $\delta$-peaks of the hard-core bosons
are located.}
\label{fig:x06}
\end{figure}

In order to make the effects more sizeable we pass to larger values of 
$x$ in Fig.\ \ref{fig:x06} displaying the results for
$x=0.6$. Note the changes of scale on the axes relative to Fig.\
\ref{fig:x02}. Still, in the panels of Fig.\ \ref{fig:x06} it is clear 
that the main broadening  of the line shapes is due to the hard-core repulsion. 
This is especially true for larger temperatures where the broadening is rather large
growing exponentially $\propto \exp(-\beta\Delta)$ 
\cite{essle08,james08,essle09,goetz10,fause14,fause15}.
 with temperature
in the low-temperature regime. Noticeable in Fig.\ \ref{fig:x06} 
is the tiny effects of the pure mean-field corrections to the
hard-core calculation. Simple frequency-independent Hartree-
and Fock-corrections only influence the dispersion a bit and shift
the positions of the line shapes. The resulting curves are very close
to the pure hard-core line shapes, especially at higher temperatures
where the lines are rather broad anyway.

The main observation is that the inclusion of the additional interactions
enhance the broadening. Thus the peaks become lower
because they become broader. This is rather striking at the lower temperature
($T=0.3J$) where the peak are still very high and prominent. At the higher 
temperature ($T=0.6J$) the effect is less obvious because the peak width
due to hard-core repulsion is already large.

A second noticeable effect of the additional interactions is a shift in the
peaks. While the additional broadening was plausibly expected because
the additional interactions open additional decay channels the shifts
come as a surprise. The gap mode is shifted to lower energies while the
maximum mode is shifted to higher energies. Thus, these shifts counteract 
the tendency induced by the hard-core repulsion
of band narrowing, i.e., the lower modes are shifted to higher energies
and vice-versa. At low temperatures, the band narrowing is even inverted
because the gap mode stays at its $T=0$ while the maximum mode moves
a bit upward. At higher temperatures, however, the main effect
remains an upward shift of the gap mode although it is slightly
reduced by the effects of the additional interactions.

\begin{figure}[htb]
\centering
\includegraphics[width=1.0\columnwidth]{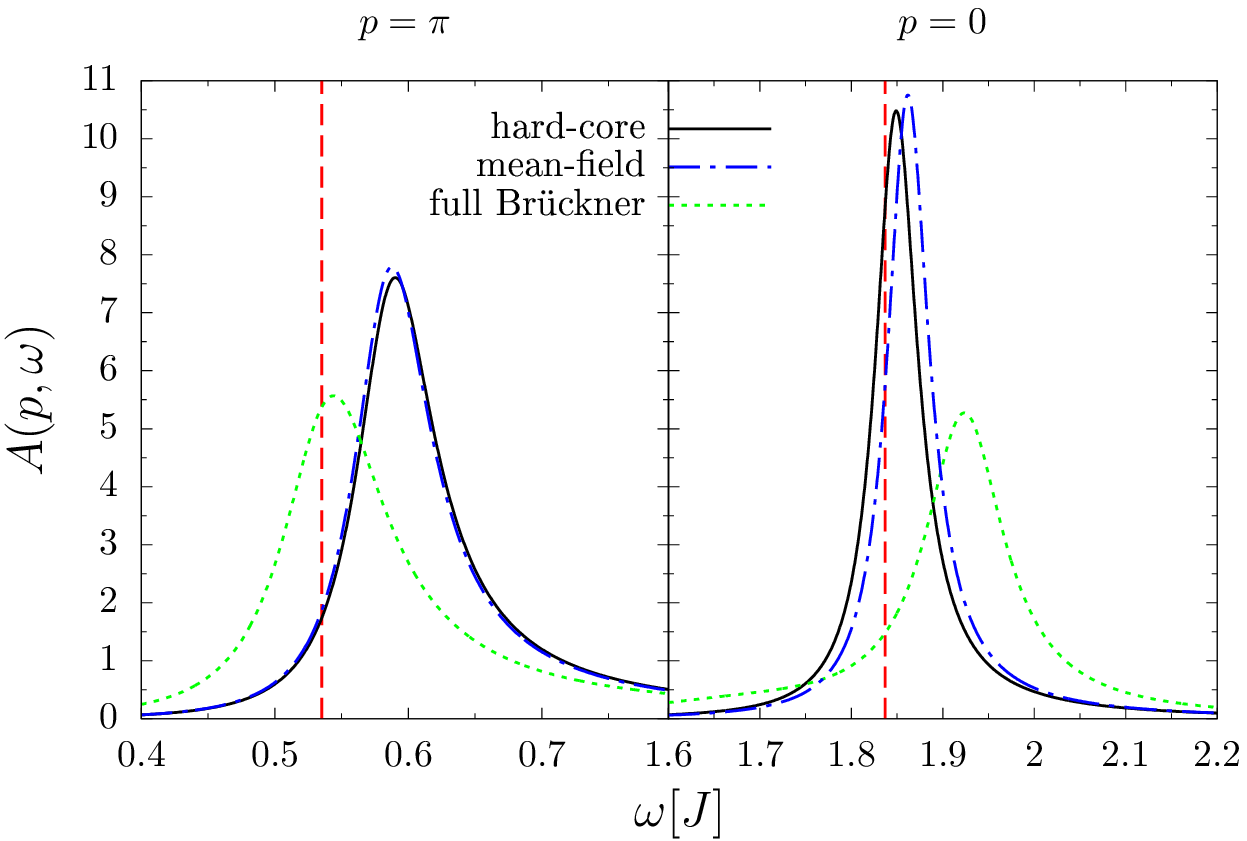}
\includegraphics[width=1.0\columnwidth]{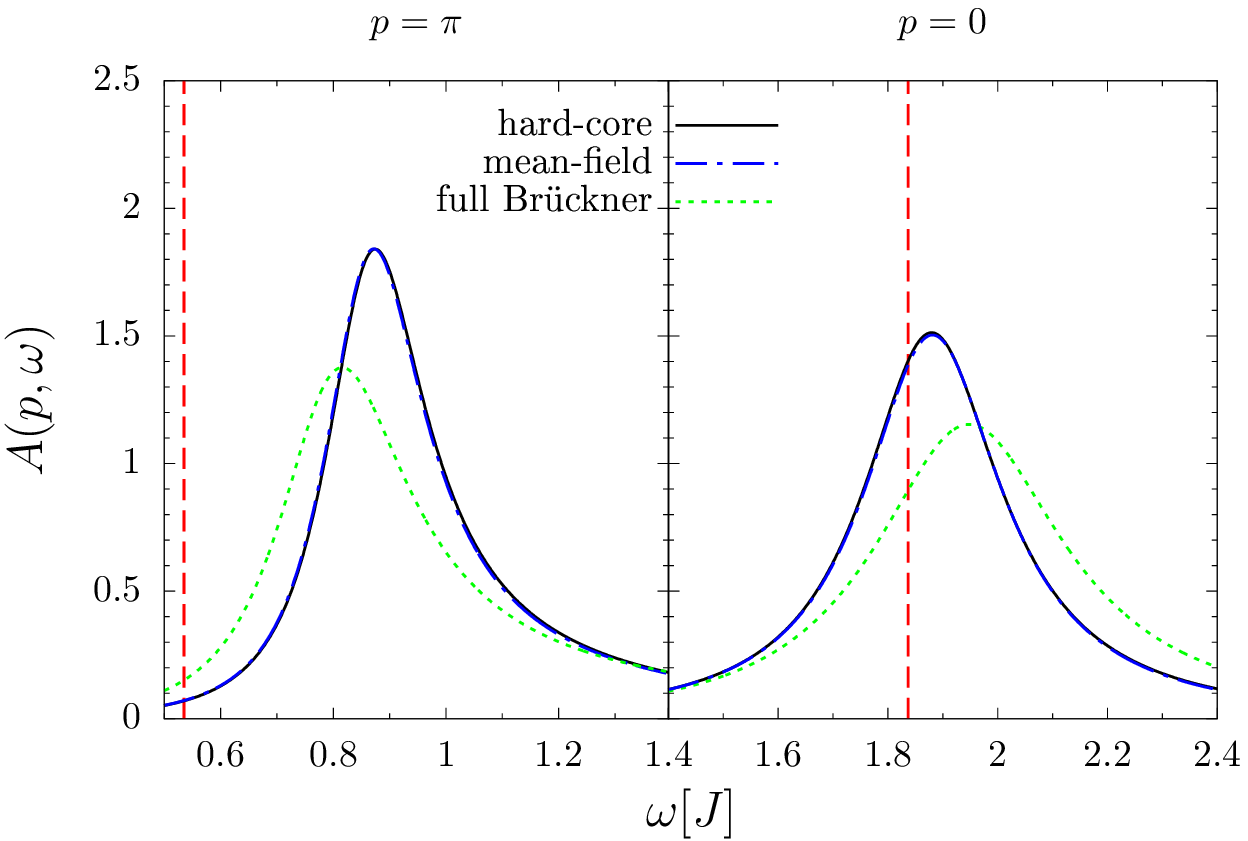}
\caption[Spectral functions  for $x=0.8$]{Spectral functions of 
for $x=0.8$ at momentum $p=\pi$ (left panels, gap mode) and $p=0$ 
(right panels). The temperature is  $T=0.3J$ for the
upper panels and $T=0.6J$ for the lower ones. The vertical dashed lines
show where the $T=0$ $\delta$-peaks of the hard-core bosons
are located.}
\label{fig:x08}
\end{figure}

\begin{figure}[htb]
\centering
\includegraphics[width=1.0\columnwidth]{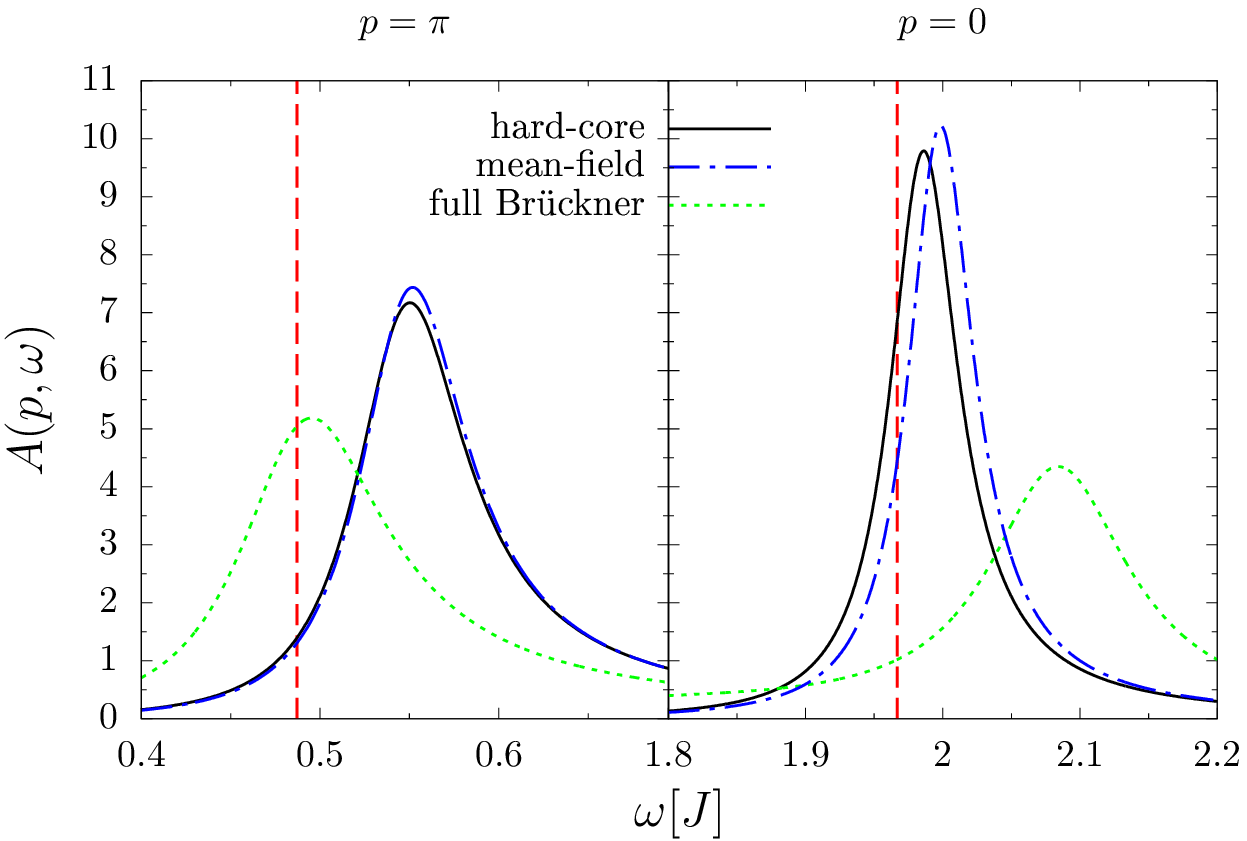}
\includegraphics[width=1.0\columnwidth]{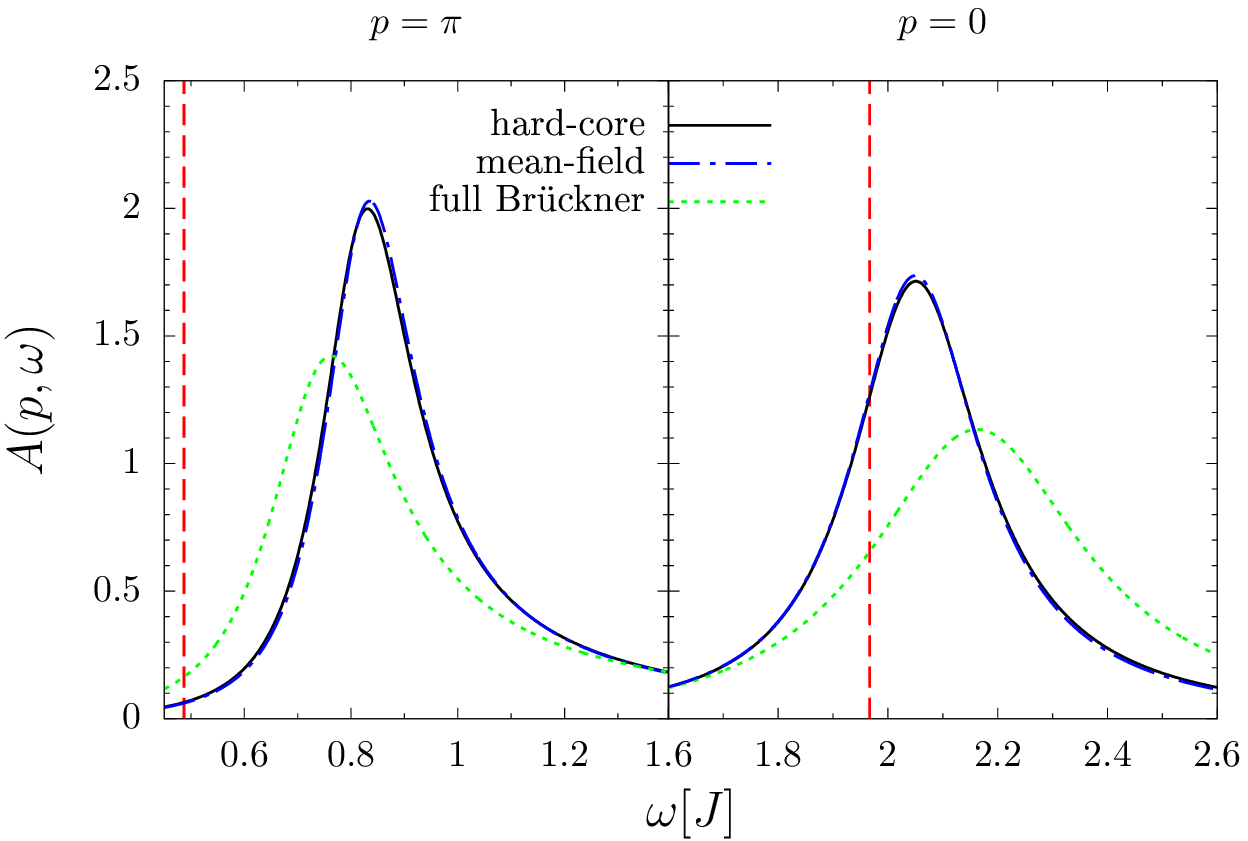}
\caption[Spectral functions  for $x=0.8$]{Spectral functions of 
for $x=1.0$ at momentum $p=\pi$ (left panels, gap mode) and $p=0$ 
(right panels). The temperature is  $T=0.3J$ for the
upper panels and $T=0.6J$ for the lower ones. The vertical dashed lines
show where the $T=0$ $\delta$-peaks of the hard-core bosons
are located.}
\label{fig:x10}
\end{figure}

These observations become more pronounced the stronger
the additional interactions are. This is corroborated by data for
increasing values of $x$ as depicted in Fig.\ \ref{fig:x08}
 and Fig.\ \ref{fig:x10}. Roughly, the broadening and the 
shifts increase with increasing $x$, but the effect is not proportional
to $x$; the increase is less than linear. We attribute
this behavior to the fact that the dominating effect in
broadening and shift still is engendered by the hard-core repulsion
which is the same in all three cases. Moreover, the band width also
increases with $x$ which limits the relative strength of the additional
interactions. 

At higher temperatures, the strong broadening
induced by the hard-core repulsion smears out the line shapes
so that the effects of additional interactions become less and less
important. This is reasonable because in the limit of infinite temperature 
only the size of the local Hilbert space matters for the dynamics of the system. 
Although we are technically working with bosons having infinite-size 
local Hilbert space the hard-core constraint implemented in the Br\"uckner approach 
prevents double and higher particle number occupation. Thus, the difference between 
the pure hard-core calculation and the calculation including additional interactions 
decreases for higher temperatures. 

These observations explain why already the hard-core repulsion 
describes experimental data very well \cite{klyus16,fause16}.
Interestingly, the shifts stemming from the additional interactions 
will improve the agreement between the diagrammatic approach and 
the peak positions computed numerically by density-matrix renormalization,
see Supplemental Material of Ref.\ \onlinecite{klyus16}.
We note that the band narrowing \cite{ cavad00b, ruegg05,mikes06,essle08,james08,essle09,goetz10, tenna12, quint12}
is reduced at high temperatures and even inverted at lower temperatures.
This calls for comprehensive further studies in theory and in experiment.

\section{Conclusions}
\label{sec.conclusion}

The goal of this article was to study how additional interactions (besides the 
hard-core repulsion) in generic hard-core
bosonic systems affect the dynamical correlations at finite temperature. 
To this end, we had to make methodical progress because the diagrammatic 
Br\"uckner approach formulated so far in solid state physics did not
include all interactions, but only the infinite on-site repulsion.

We extended the diagrammatic Br\"uckner approach by including the complete interaction 
in the summation of all  ladder diagrams. The solution of the 
extended Bethe-Salpeter equation was possible 
by introducing a scattering matrix which took over the role of the scattering amplitude
in the previous case of the pure on-site repulsion.
The result describes all possible iterated scattering processes between two given
elementary excitations. This allowed us to carry out the intricate limiting procedure
$U\to\infty$ analytically. In this way, it was possible to calculate the single-particle 
self-energy correctly in first order in the expansion parameter $\exp(- \beta \Delta)$.
Thus, we perform a systematic expansion valid for low temperatures.
The contributions to the self-energy stemming from the virtual antibound state as well as 
from the low-energy sector due to the additional interactions were derived explicitly.

In order to demonstrate how the method works and to illustrate the importance
of the additional interactions we applied the method to a well-understood
system with established hard-core bosonic excitations, namely the Heisenberg spin ladder.
Here, the elementary excitation are excited spin dimers on the rungs
of the ladder including their dressing on the rungs in the vicinity. Due to 
their total spin $S=1$ they are called triplons and realize hard-core bosons
with three flavors corresponding to the three states of a triplet. The
effective model at zero temperature expressed in triplon creation and annihilation operators
is available for instance by continuous unitary transformations.

At finite temperatures, we compared the line shapes resulting from 
pure hard-core scattering, from hard-core scattering complemented by mean-field
corrections, and from the full Br\"uckner approach. We showed that the dominating
effect is the broadening of the $T=0$ $\delta$-peak by the hard-core scattering
at finite temperature. The mean-field corrections provide only small modifications.
They induce small, almost negligible shifts in the peak positions.
The additional interactions, however, have noticeable effects. 
They induce signatures of bound and antibound states in the spectral functions
of the scattering matrix. These are not directly detectable, but the
effects on the spectral functions are clear.
The additional interactions broaden the lines further since they
provide additional decay channels. Concomitantly, they induce certain shifts
in the peak positions. Interestingly, these counteract the shifts
induced by the hard-core repulsion. The latter imply
a certain band narrowing pushing low-lying modes upwards in energy 
and high-lying ones downward. So the inclusion of the additional interactions
reduce this effect of band narrowing and may even invert it
for low temperatures. These additional shifts are likely to improve the
agreement with experimental observations further.

As an outlook, we point out that besides the single-particle response,
multi-particle response can play a significant role in 
real experiments. The leading effect is given by the broadening 
induced in the single-particle propagators which carries over to the 
multi-particle response. Villain pointed out \cite{VILLAIN19751} that for thermal 
excitations intraband transitions with arbitrary small
energy differences are possible leading to a low-energy
response at $\omega\approx 0$. 
This mechanism was discussed and analyzed by Essler and co-workers for the
alternating spin chain and the spin ladder in the limit of strong coupling on the dimers and rungs, respectively
\cite{james08,goetz10}.
At zero temperature, the intraband response completely
vanishes because no triplons are thermally excited.
Once the temperature is finite, the quasi-particle band is populated
and intraband transitions become possible, inducing
a finite spectral weight of the low-energy response. In
terms of the effective model, such intraband transitions
can appear if the corresponding observable includes
terms proportional to $t^\dagger t$ which is generically
the case. Thus the intraband transitions
can be interpreted as the propagation of a quasi-particle
and an annihilated thermal quasi-particle. Therefore, in first
order in $\exp(-\beta\Delta)$, the response at low energies
can be calculated by the convolution of single-particle
propagators obtained within the Br\"uckner approach.
A quantitative discussion is beyond the scope of the present
paper, but subject of future research.

Summarizing, the main effect results from the hard-core repulsion
as was to be expected from the size of the matrix elements (here $U\to\infty$).
But for quantitative analyses, the effect of additional interactions is
indeed very important and cannot be neglected. The attempt to take
them into account on the mean-field level does not capture the
relevant size of the shifts and fails to capture the additional broadening.
These insights have only become possible due to the methodical extension 
of the Br\"uckner approach from pure on-site interaction to general
interactions of finite range. The so far scalar geometric series at the
basis of the solution of the Bethe-Salpeter equation had to be promoted
to a matrix-valued geometric series. Clearly, the application to a wide
range of gapped systems is possible. In particular, we emphasize that
the Br\"uckner approach can be applied in any dimension.

\bibliographystyle{apsrev}

\appendix

\begin{widetext}

\section{Bosons with multiple flavors}
\label{app.c}

In case of multi-flavored bosons, the most general two-particle interaction is
 parameterized by
\begin{align}
V &= \sum\limits_{j} \sum\limits_{d_1, d_2, d_3} \sum\limits_{\alpha, \beta, \gamma, \delta} 
V^{\alpha \beta \gamma \delta}_{d_1,d_2,d_3} 
b^\dagger_{j,\alpha} b^\dagger_{j+d_1,\beta} b^{\phantom{\dagger]}}_{j+d_2,\gamma} 
b^{\phantom{\dagger}}_{j+d_2+d_3,\delta},
\end{align}
where $d_1,d_2,d_3 \in \mathbb{Z}$ and $d_1 \neq 0$ and $d_3 \neq 0$ holds due to the 
hard-core constraint. The flavor indices are given by $\alpha, \beta, \gamma, \delta$.
In momentum space the interaction reads
\begin{align}
V=\frac{1}{N} \sum\limits_{p,k,q} \sum\limits_{d_1, d_2, d_3} 
\sum\limits_{\alpha, \beta, \gamma, \delta} 
V^{\alpha \beta \gamma \delta}_{d_1,d_2,d_3} 
b^\dagger_{p+k,\alpha} b^\dagger_{-k,\beta} b^{\phantom{\dagger]}}_{p+q,\gamma} b^{\phantom{\dagger}}_{-q,\delta} 
\cdot e^{-i d_2 p} e^{-i d_1 k} e^{ i d_3 q}
\end{align}
The interaction vertex in case of flavored bosons reads
\begin{align} \label{eq.new_interaction3_flavor}
V^{\alpha \beta \gamma \delta}(p,k,q) = \frac{U}{N} + \frac{1}{N} \sum\limits_{d_1, d_2, d_3} 
V^{\alpha \beta \gamma \delta}_{d_1,d_2,d_3} e^{-i d_2 p} e^{-i d_1 k} e^{i d_3 q}.
\end{align}
The corresponding diagram is represented in Fig.\ \ref{image:new_interaction3}.

\begin{figure}[ht]
\centering
\includegraphics[width=0.15\textwidth]{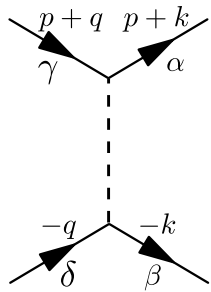}
\caption[Diagrammatic representation of the interaction for multi-flavored
bosons]{Diagrammatic 
representation of Eq.\ \eqref{eq.new_interaction3_flavor} for multi-flavored bosons.}
\label{image:new_interaction3}
\end{figure}

In the Heisenberg ladder, two kinds of interactions of the hard-core
triplons are present: 
$V^{\alpha \alpha \gamma \gamma}$ for interactions with the same flavor for the 
ingoing triplons and the same flavor for the outgoing triplons and 
$V^{\alpha \gamma \alpha \gamma}$ for the interaction of bosons with different flavors. 
Hence, we will focus our analysis on this special case.

The main change in comparison to the single-flavor case is that the scattering matrix 
also acquires flavor indices, which represent super indices. As a result, the matrix 
dimension for $\uuline{\Gamma}$ scales with the squared number of flavors $N_\mathrm{f}^2$.
First, we solve the Bethe-Salpeter equation for the case $V^{\alpha \alpha \gamma \gamma}$
\begin{align}
\uuline{\Gamma}^{\alpha \alpha \gamma \gamma}(P) &= \left(  
{\uuline{\Gamma}_{0}^{\alpha \alpha \beta \beta}}^{-1}(p) + \uuline{N}(P) \right)^{-1}.
\end{align}
Second, for $V^{\alpha \gamma \alpha \gamma}$
\begin{align}
\uuline{\Gamma}^{\alpha \gamma \alpha \gamma}(P) &= \left(  
{\uuline{\Gamma}_{0}^{\alpha \gamma \alpha \gamma}}^{-1}(p) + 
\uuline{N}(P)   \right)^{-1}.
\end{align}
Here, we assumed that the dispersion of the different flavors is the same, i.e., 
$G^\alpha(P) = G(P)$ independent of $\alpha$. This is the case if the SU(2) invariance 
is not broken in the Hamiltonian. 

From the matrix $\uuline{\Gamma}^{\alpha \alpha \gamma \gamma}(P)$ we only need the 
flavor-diagonal  parts $\alpha = \gamma$ due to the structure of the diagrams in Fig.\ 
\ref{image:Ladder5_example}. But the non-diagonal parts of $V^{\alpha \alpha \gamma \gamma}$
 mix with the diagonal elements in the diagonalization of the scattering matrix.

Finally, we can calculate the Hartree- and Fock-like diagrams for the self-energy. Note 
that for $\uuline{\Gamma}_{\alpha \gamma \alpha \gamma}(P)$ only the Hartree-like 
diagrams contribute while for the $\uuline{\Gamma}^{\alpha \alpha \alpha \alpha}(P)$ 
both, the Hartree- and the Fock-like diagrams contribute. 
Besides the distinction between the two different $\uuline{\Gamma}$ matrices, the 
calculation of the self-energy remains unchanged.

\section{Matrix perturbation theory for $X_1$ and $X_2$}
\label{app.b}
 
In order to be able to compute the proper limit $U\to\infty$ for the
self-energy we need the first order corrections  in $1/U$ 
of the eigen value $\Delta^{(1)}\lambda_1$ and the eigen state 
$\Delta^{(1)} \mathbf{e}_1$ as well as the second
order correction of the eigen value $\Delta^{(2)} \lambda_1$,
see Eq.\ \eqref{eq.eigenpair}.
From standard perturbation theory we obtain
\begin{subequations}
\begin{align}
\Delta^{(1)} \lambda_1 &= \mathbf{e}_1^\dagger X_1 \mathbf{e}_1 = 
\frac{\rho_0^{11}(p)}{\bar{\omega}} + 1 \\
\Delta^{(1)} \mathbf{e}_1 &= \sum\limits_{i \neq 1} 
\underbrace{\frac{\mathbf{e}_i^\dagger X_1 
\mathbf{e}_1}{\lambda_i}}_{\frac{\mathbf{e}_i^\dagger \underline{\underline{\rho}}_0(p) 
\mathbf{e}_1}{\lambda_i \bar{\omega}}} \mathbf{e}_i 
\\
\Delta^{(2)} \lambda_1 &= \sum\limits_{i \neq 1} \underbrace{\frac{ 
\left| \mathbf{e}_i^\dagger X_1 \mathbf{e}_1 \right|^2}{\lambda_i} }_{\frac{ 
\left| \mathbf{e}_i^\dagger \underline{\underline{\rho}}_0(p)  
\mathbf{e}_1 \right|^2}{\lambda_i \bar{\omega}^2} } + 
\underbrace{ \mathbf{e}^\dagger_1 X_2 \mathbf{e}_1 }_{\frac{\rho_1^{11}(p)}{\bar{\omega}^2}}.
\end{align}
\end{subequations}
Note, that one does not need the second order corrections of the eigen vector 
in the following, since it does not contribute in the limit $U\to\infty$.

We introduce the abbreviations
\begin{align}
W_i(p) &:=\frac{\mathbf{e}_i^\dagger \underline{\underline{\rho}}_0(p) 
\mathbf{e}_1}{\lambda_i} , \quad 
V_i(p) :=\frac{\left|\mathbf{e}_i^\dagger \underline{\underline{\rho}}_0(p) 
\mathbf{e}_1 \right|^2}{\lambda_i} .
\end{align}
To find the pole of the antibound state, the matrix $\uuline{\Gamma}(P)$ must be singular 
for high frequencies of the order of $U$, i.e., the first eigen value must vanish 
as function of $\bar{\omega}$
\begin{subequations}
\begin{align}
0 &\overset{!}{=} \lambda_1 + \frac{1}{U} \Delta^{(1)} \lambda_1 + \frac{1}{U^2}\Delta^{(2)}
 \lambda_2  
\\
0 &= \frac{\rho_0^{11}(p)}{\bar{\omega}} + 1 + \frac{1}{U} \sum\limits_{i \neq 1} V_i(p) 
\frac{1}{\bar{\omega}^2} + \frac{1}{U} \frac{\rho_1^{11}(p)}{\bar{\omega}^2} 
\\
\Rightarrow \bar{\omega} &= -\rho_0^{11} + \frac{1}{U} \left(  \sum\limits_{i \neq 1} 
\frac{V_i(p)}{\rho_0^{11}(p)} + \frac{\rho_1^{11}(p)}{\rho_0^{11}(p)} \right) + 
\mathcal{O} \left( \frac{1}{U^2} \right).
\end{align}
\end{subequations}
Hence the pole occurs at the frequency
\begin{align}
\omega_U(p) = U \bar{\omega} =  -U\rho_0^{11}(p) +  \left(  \sum\limits_{i \neq 1} 
\frac{V_i(p)}{\rho_0^{11}(p)} + \frac{\rho_1^{11}(p)}{\rho_0^{11}(p)} \right) + 
\mathcal{O} \left( \frac{1}{U^1} \right)
\end{align}
Expanding $\lambda_1(\omega)$ around $\omega = \omega_U$ 
the eigen value is approximated by
\begin{align}
\lambda_1(\omega) \approx  \left( \omega - \omega_U(p) \right) 
\frac{-1}{U^2 \rho_0^{11}(p)} + \mathcal{O} \left( \frac{1}{U^2} \right) .
\end{align}

To obtain the scattering matrix at high energies 
the correction to the eigen vector must also be considered
\begin{align}
\mathbf{e}_1' = \mathbf{e}_1 + \frac{1}{U \bar{w}} \Delta^{(1)} \mathbf{e}_1 = 
\mathbf{e}_1 - \frac{1}{U} \sum\limits_{i \neq 1} \frac{W_i(p)}{\rho_0^{11}(p)} \mathbf{e}_i .
\end{align}
Thus we can approximate the scattering matrix for large energies according to 
\begin{subequations}
\begin{align}
\left(\uuline{\Gamma}_{0}^{-1} + \uuline{N}(P)\right)^{-1} &\approx 
\frac{-U^2 \rho_0^{11}(p)}{\left( \omega - \omega_U(p) \right)} 
\mathbf{e}_1'(p) \mathbf{e}_1'^\dagger(p) 
\\
&= \frac{-U^2 \rho_0^{11}(p)}{\left( \omega - \omega_U(p) \right)} 
\left[ \mathbf{e}_1 \mathbf{e}_1^\dagger - \frac{1}{U} \left( \mathbf{e}_1 
\sum\limits_{i \neq 1} \frac{W_i^*(p)}{\rho_0^{11}(p)} \mathbf{e}_i^\dagger + 
\sum\limits_{i \neq 1} \mathbf{e}_i \frac{W_i(p)}{\rho_0^{11}(p)} \mathbf{e}_1^\dagger \right) 
 +  \mathcal{O} \left( \frac{1}{U^2} \right) \right]  
\end{align}
\end{subequations}

To obtain the correct scalar contribution to the self-energy, we need to calculate the 
bilinear form in Eq.\ \eqref{eq.Ansatz_Gamma}. To shorten the notation, we first introduce
\begin{align}
 f_i(q) :=  \mathbf{e}_i^\dagger \mathbf{f}(q).
\end{align}
Next, we compute the imaginary part of the scattering amplitude for large energy 
including order $1/U$
\begin{equation}
\begin{aligned}
\frac{-\mathrm{Im}}{\pi N} \lim_{i \omega_p \rightarrow \omega + i0+} 
\mathbf{f}^\dagger(k) \uuline{\Gamma}(P) \mathbf{f}(q)
&= \frac{-U^2 \rho_0^{11}(p)}{N\beta} \delta \left(\omega - \omega_U(p) \right) 
\left[ 1- \frac{1}{U}  \sum_{j \neq 1} 
\frac{1}{\rho_0^{11}(p)}\left(W_j^*(p) f_j(q) + W_j(p) f_j(k)^* \right) \right] .
\end{aligned}
\end{equation}
Finally, this expression for the scattering amplitude is used to
compute the self-energy contributions given in the main text.
\end{widetext}
\end{document}